\documentclass[twocolumn,showpacs,preprintnumbers,amsmath,amssymb]{revtex4}

\usepackage{graphicx}
\usepackage{dcolumn}
\usepackage{bm}

\begin{document}

\title{Electron spin evolution induced by interaction with nuclei in
 a quantum dot}

\author{Alexander Khaetskii$^1$, Daniel Loss$^1$, and Leonid Glazman$^2$}
\address{$^{1}$
Department of Physics and Astronomy,
University of Basel,
Klingelbergstrasse 82,
CH-4056 Basel, Switzerland}

\address{$^{2}$
Theoretical Physics Institute, University of Minnesota, Minneapolis, MN
55455, USA.}

\begin{abstract}
We study the decoherence of a single electron spin in an isolated
quantum dot induced by hyperfine interaction with nuclei for times
smaller than the nuclear spin relaxation time. 
The decay is caused by the spatial variation of the electron envelope
wave function within the dot, leading to a non-uniform hyperfine coupling $A$.  
We show that the usual treatment of the problem based on the Markovian approximation is impossible
because the correlation time for the nuclear magnetic field seen by the electron spin is itself determined by the flip-flop processes.  
 The decay  of the electron spin correlation function is not exponential but rather  power (inverse logarithm) law-like. 
For polarized nuclei we find an exact solution  and show that
the precession amplitude and the  decay behavior  can be tuned by the magnetic field. The decay time is given by $\hbar N/A$, where $N$ is the number of  nuclei
inside the dot.  
The amplitude of precession, reached as a result of
the decay, is finite.
We show that there is a striking difference
between the decoherence time for a single dot and the dephasing time
for an ensemble of dots.
\end{abstract}

\maketitle

\section{Introduction} 
\label{intro}

The spin dynamics of  electrons in semiconducting
nanostructures has become of central interest in recent years \cite{Spintronics,Wolf}.
The controlled manipulation of spin, and in particular of its phase, 
is the primary prerequisite needed for novel applications in conventional
computer hardware as well as in quantum information processing.
It is thus desirable to understand the  mechanisms
which limit the spin phase coherence of electrons, in particular in 
GaAs semiconductors, which have been shown \cite{Awschalom} to exhibit unusually
long spin decoherence times $T_2$ exceeding 100 ns.
Since in GaAs  each nucleus carries spin, the hyperfine
interaction between electron and nuclear spins is unavoidable, and it
is therefore important to underdffect on the electron
spin dynamics \cite{Salis}.  This is particularly so
for  electrons which are confined
to a closed system such as a quantum dot with a spin 1/2 ground state,
since, besides fundamental interest, these systems are promising candidates for 
scalable spin
qubits \cite{Loss98}. 
For recent work on spin relaxation (characterized by $T_1$ times)
in GaAs
nanostructures we refer to Refs.\cite{Khaetskii,Nazarov,Lyanda}.

Motivated by this we
investigate in the following the spin dynamics of a single electron
confined to a quantum
dot in the presence of nuclear spins (see Fig. 1)\cite{We}. We treat  the case of unpolarized
nuclei perturbatively, while for the fully polarized case we present
an exact solution for the spin dynamics and show that the decay
is non-exponential and can be strongly influenced by external magnetic fields.
 We
use the term "decoherence" to describe the case with a single dot,  
and the term "dephasing" for an ensemble
of  dots \cite{ensemble}. 
The typical
fluctuating nuclear magnetic field  seen by the electron spin 
via the hyperfine
interaction   is of the order of\cite{Dyakonov}
$\sim A/(\sqrt{N} g \mu_B)$, with an associated electron precession
frequency  $\omega_N \simeq A/\sqrt{N}$,
where $A$ is a hyperfine constant, $g$  the
electron g-factor, $\mu_B$  the Bohr magneton, and $N$ the number of nuclei inside the dot. 
 For a  typical dot size the electron wave function covers approximately $N=10^5$
nuclei, then this field is of the order of 100 Gauss in a GaAs quantum dot.
The nuclei in turn interact with each other via dipolar
interaction, which does not conserve the total nuclear spin
and thus leads to a change of a given nuclear spin configuration within the time
$T_{n2} \approx 10^{-4} s $, which is just the period of precession of a nuclear
spin in the local magnetic field  generated by its neighbours.

We note that  there are two    different regimes of interest,  
depending on the parameter $\omega_N \tau_c$, where
$\tau_c$ is the correlation time of the  nuclear magnetic field ${\bf H}_N(t)$ seen by the electron spin 
via the hyperfine interaction.
  The  simplest case is the perturbative regime
$\omega_N \tau_c \ll 1$,  
characterized by dynamical narrowing: 
different random nuclear configurations  change quickly in time, and, as a
result, the spin dynamics is diffusive with a
dephasing time $ \simeq 1/ (\omega_N^2 \tau_c)$.  
This case is realized, for example, for a system of quantum dots (or shallow donors)
when the hopping rate $1/\tau_c$ of the electron between  neighbouring dots is high. The problem of electron spin relaxation for the case of electron hopping between shallow donors in GaAs was studied in Ref. \onlinecite{Dyak}. 
Using a perturbative approach, we easily obtain the following formulas for the
 longitudinal ($T_1$) and transverse ($T_2$) spin relaxation times \cite{Dyak,Abragam}:
\begin{widetext}
\begin{eqnarray}
\frac{1}{T_1}= \frac{(g\mu_B)^2}{\hbar^2} \frac{1}{2} \int_{-\infty}^{+\infty} d\tau 
\exp(-i\omega_z \tau) Re [< H_{Nx}(0) H_{Nx}(\tau)> + < H_{Ny}(0) H_{Ny}(\tau)>], \nonumber \\ 
\frac{1}{T_2}= \frac{1}{2T_1} + \frac{(g\mu_B)^2}{\hbar^2} \frac{1}{2} 
\int_{-\infty}^{+\infty} d\tau < H_{Nz}(0) H_{Nz}(\tau)>, 
\label{usual}
\end{eqnarray}
\end{widetext}
 
where $\omega_z =g\mu_B B/\hbar$ is the Larmor spin precession frequency in the external magnetic  field $B$ directed along the z-axis, and $<...>$ means the ensemble average of the fluctuating nuclear magnetic field correlators. (In the limit $\omega_z \tau_c \ll 1$ we obtain from the above formulas  the well known result $T_1 = T_2$, where we took into account the fact that the fluctuating nuclear field is isotropic.)  

A more difficult situation arises when $\omega_N \tau_c \gg 1$, which requires
a nonperturbative approach. It is this regime that we will
consider in this paper, i.e.  the electron is localized in a quantum dot, and  the
correlation time is due to the internal nuclear spin dynamics, i.e., $\tau_c=T_{n2}$,
giving $\omega_N \tau_c=10^4$. 
Thus, no usual treatment  and no Markov approximation are possible. 
In particular, the perturbative formulas for $T_1, T_2$ (see Eqs.(\ref{usual})) are not applicable here. 
Next, we need to address the important issue of averaging over different nuclear 
spin configurations in a single dot. Without internal nuclear 
spin dynamics, i.e. $T_{n2}\rightarrow  \infty$, no averaging is indicated. However,
each flip-flop process  (due to hyperfine interaction)
creates a different
nuclear spin configuration, and because of the spatial variation of the hyperfine
coupling constants inside the dot, this
leads  to a different random value of the  nuclear  field seen by the
electron spin and thus to its decoherence. Below we will find
that this decoherence is non-exponential, but still we can indicate the
characteristic time given by \cite{ensemble} $(A/\hbar N)^{-1}$. Moreover, we shall find
that $T_{n2}\gg (A/\hbar N)^{-1}$, and thus
 still no averaging over the nuclear configurations is indicated (and
dipolar interactions will be neglected henceforth).
To underline the importance of this point, we will  contrast below
the unaveraged correlator with its average.

\section{Unpolarized nuclei.} 
\label{unpolarized}

We consider a single electron in an orbital ground state of a quantum dot. Its spin ${\bf S}$
couples to an external magnetic field ${\bf B}$ and to
nuclear spins $\{{\bf I}^i\}$ via hyperfine contact interaction, described
by  the Hamiltonian 
\begin{equation}
\hat {\cal H}= g\mu_B {\bf S} \cdot {\bf B} + {\bf S} \cdot {\bf h}_N\, , \/
 \, {\bf h}_N =  \sum_i A_i {\bf I}^i =  g\mu_B {\bf H}_N \, , 
\label{Ham}
\end{equation}   
where ${\bf H}_N$ is the nuclear field. Note that the sum in Eq.(\ref{Ham}) runs over the entire space.
The coupling constant with the i-th nucleus, $A_i =A v_0 | \Psi({\bf r}_i) |
^2$, contains the  electron  envelope
wave function $\Psi({\bf r}_i)$ at the nuclear site ${\bf r}_i$, and $v_0$ is the
volume of the crystal cell.  
For simplicity we consider nuclear spin 1/2. Neglecting dipolar  
interactions between the
nuclei, we can  consider only  some particular nuclear configuration,
described in the $\hat I_z^i$ eigenbasis as
$|\{I_z^i\}>$, with $I_z^i=\pm 1/2$.\cite{general,conserv}
Moreover, we
assume an unpolarized configuration with a typical net nuclear magnetic
field $A/(\sqrt{N}g\mu_B)$, being much less than 
$ A/(g\mu_B)$ (fully polarized case), for the precise definition of $N$ see below. 
 We study  the decay  of the electron spin from its initial ($t=0$)
$\hat S_z$-eigenstate $| \Uparrow >$.
For this we evaluate the correlator  
\begin{equation}
C_n(t)=<n \mid \delta \hat  S_z(t) \hat S_z \mid n>, 
\label{corr}
\end{equation}
where $<\mid...\mid >$ means the diagonal matrix  element. 
Here $\delta \hat S_z(t) =  \hat S_z(t)- \hat S_z$, and 
$\hat S_z(t) = \exp(it \hat {\cal H}) \hat S_z \exp(-it \hat {\cal H})$.
This correlator is proportional to $<n|\hat S_z(t)- \hat S_z(0)|n>$.
Since at $t=0$ the total (electron and nuclear)
state $|n>=|\Uparrow, \{I_z^i\}>$ is an eigenstate of $ \hat {\cal H}_0 = \hat S_z \hat 
h_{Nz}$ (with eigenergy $\epsilon_n$),   we can expand in the perturbation    $\hat
V= (1/2)(\hat S_+ \hat h_{N-} + 
\hat S_- \hat h_{N+})$ (with $\hat {\cal H}= \hat {\cal H}_0 + \hat V$).
Introducing the usual time evolution operator $\hat U(t) = \hat T \exp(- i \int_0^t dt_1 \hat V (t_1)) $, with $\hat T$ being the time-ordering operator, we get for the correlator 
$ <n\mid \hat U^{\dagger}(t) \hat S_z(t) \hat U(t) \hat S_z(0) \mid n >$, where  the 
time dependence of all operators is due to  the $ \hat {\cal H}_0$ 
Hamiltonian. Then 
we obtain in  leading order in $\hat V$,
\begin{widetext}
\begin{equation} 
C_n(t) =  \sum_k \frac{|V_{nk}|^2}{\omega_{nk}^2} (
\cos (\omega_{nk} t)  -1 ) \approx \frac{1}{[\epsilon_z +(h_z)_n]^2} \sum_k \frac{A_k^2}{8}[
\cos (\epsilon_z +(h_z)_n + A_k/2)t  -1 ]  \, ,
\label{4}
\end{equation} 
\end{widetext}
where $V_{nk}=<n|\hat V|k>$ is the matrix element 
between initial state
$n=\Uparrow, \{..., I^k_z = -1/2,... \}$  and intermediate state $k= \Downarrow, 
\{...., I^k_z = +1/2,... \} $, 
 and $\omega_{nk}= \epsilon_n -\epsilon_k$, $\epsilon_z = g\mu_B B_z$. 
We have used that $|V_{nk}|^2  = 
A_k^2 <n| 1/2- \hat I^k_z | n>/4$, and $\omega_{nk} = \epsilon_z +(h_z)_n + A_k/2$, where
$(h_z)_n = <n | \hat h_{Nz} | n>$, and the fact that for the typical nuclear
configuration  $(h_z)_n^2 \simeq \omega_N^2 \gg A_k^2$.
 Since $N\gg 1$, we replace the sums over $k$ (which run over the
entire space)  by integrals, i.e. $\sum_k f_k =(\int d^3 r/v_0) f({\bf r}) + o(1/N)$. 
Then we have
\begin{widetext}
\begin{equation}
C_n(t) 
\simeq - \frac{A^2}{8\pi N (\epsilon_z+(h_z)_n)^2}  \left [\frac{I_0}{2}- I_1 (\tau)\cos (\epsilon_z +(h_z)_n) t  + I_2(\tau) \sin (\epsilon_z +(h_z)_n )t \right], 
\label{cor1}
\end{equation}
\end{widetext}
where 
\begin{equation}
I_0 =\int_{-\infty}^{+\infty} dz \chi_0^4(z), \/
I_{1,2}(\tau) =\int_{-\infty}^{+\infty} dz \chi_0^4(z)F_{1,2}(\tau \chi_0^2(z)), 
\label{I}
\end{equation}
and
\begin{equation}
F_1(\eta)= \frac{\sin \eta}{\eta} + \frac{\cos \eta -1}{\eta^2}; \/
F_2(\eta)= \frac{\sin \eta}{\eta^2} - \frac{\cos \eta}{\eta}.
\label{F}
\end{equation}
Here $N = a_z a^2/v_0 \gg 1$ is the number of  nuclei inside the
dot, and
$\tau =  A t /2\pi N$.  Eq.(\ref{cor1}) was obtained   
with the use of the following expression for the electron envelope wave function:
\begin{equation}
|\Psi(\rho,z)|^2 = (1/\pi a^2 a_z) \exp(- \rho^2/a^2) \chi_0^2(z).
\label{func}
\end{equation}
Here 
$a, a_z$ are the dot sizes in the lateral and transverse (perpendicular to the
2D plane) directions, respectively,
and  the transverse wave
function  $\chi_0(z)$ is normalized, i.e. 
$\int_{-\infty}^{+\infty} dz
\chi_0^2(z)=1$.  For any analytic function $\chi_0^2(z)$ with  expansion
$\chi_0^2(z)= \chi_0^2(0)-z^2(\chi_0^2)''/2$ near its maximum, we have
$I_{1,2}(\tau \gg 1)= \pm (\chi_0^2(0)/\tau^{3/2}) 
\sqrt{\pi/(\chi_0^2)''} [\sin (\tau \chi_0^2(0))
\mp \cos(\tau \chi_0^2(0))]$.  
Thus, for this case
 we obtain the universal 
{\em power law decay}  for times
$\tau \gg 1$, i.e. $t \gg (A/N)^{-1}$:
\begin{eqnarray}  
C_n(\tau \gg 1)
\simeq  -\gamma + \frac{\tilde\gamma}{\tau^{3/2}}\sin(\tilde h_n t -\phi), \nonumber \\
\tilde h_n = \epsilon_z + (h_z)_n + A_0/2. 
 \label{cor2}
\end{eqnarray} 
Here $A_0$ is the coupling constant with the nucleus located at the center of the dot; and 
$\phi$ is a phase shift with $\phi \sim 1$. Note that for the
typical nuclear configuration the quantity 
$A^2/N(h_z)_n^2 $ is of  order unity, thus, for a weak Zeeman field  $\epsilon_z 
< \omega_N $ the part of the 
 electron spin state which decays is of the order of the initial value. 
Hence, the same holds for the spin part which survives at $\tau \gg 1$ (i.e. $\gamma \simeq \tilde\gamma \simeq 1/2$, for $\epsilon_z 
< \omega_N $). 
We see from Eq.(\ref{cor1}) that in the presence of
a large Zeeman field, $\epsilon_z
\gg \omega_N$,
the asymptotic behaviour of $C_n(t)$  is not
changed, the only difference being that 
the decaying part of the initial spin state  is small now, 
i.e. $\gamma\simeq \tilde \gamma \simeq (\omega_N/\epsilon_z)^2 \ll 1$.
For the fully polarized nuclear state  Eqs.(\ref{cor1},\ref{cor2}) should be
multiplied by 2, and $(h_z)_n$ should be replaced by $A/2$.

We note  that $C_n(t)$ in (\ref{4}) is 
quasiperiodic in $t$, and, thus, it will decay  only up to the Poincar\'{e} recurrence
time $\tau_P$.
This time
can be found from the condition that the terms  omitted when converting  sums
to  integrals  become comparable with the integral itself.
This will happen at $\tau
\simeq N$, giving $\tau_P=0.1s - 1 s$.

In  next order, $\hat V^4$, we face the problem of secular terms or
"resonances", i.e. the  corrections
will contain  zero denominators. This gives rise to linearly growing terms
$\propto\omega_N t$, even for $t
\ll (A/N)^{-1}$. In higher order  the degree of the divergence
will increase.     This means that the decay law we found can, in principle,
change after proper resummation, 
because no   small expansion parameter  exists, which, strictly speaking, would justify a perturbative
approach. Still, the result found in lowest order remains qualitatively correct in that it
shows that a non-uniform hyperfine coupling leads to a non-exponential decay of the
spin. This conclusion is confirmed  by an exactly solvable case to which we turn
next.

 \section{Polarized nuclei. Exactly solvable case.}  
\label{polarized}

In this section we
consider the exactly solvable case where the initial
 nuclear spin configuration is fully  polarized (see Fig. 2). We also allow for a magnetic field
but neglect its effect on the nuclear spins.
With the initial wave function 
\begin{equation}
\Psi_0 = |\Downarrow; \uparrow, \uparrow, \uparrow,...>
\label{0}
\end{equation}
 we can construct the {\em exact}
wave function of the system for $t>0$, 
\begin{equation}
\Psi(t)= \alpha(t) \Psi_0 + \sum_k \beta_k(t) |\Uparrow; \uparrow, \uparrow,
\downarrow_k,\uparrow,...>,
\label{function}
\end{equation} 
with normalization $|\alpha(t)|^2 + \sum_k
|\beta_k(t)|^2 =1$, and we assume that $\alpha(t=0^+) =1, \/ \alpha(t<0) =0 $. 
The second term in Eq.(\ref{function}) is an entangled coherent superposition of the states with exactly one nuclear spin flipped, and thus similar to a magnon excitation. 
The correlator $C_0$ is expressed through  $\alpha(t)$ by the formula:
\begin{equation}
C_0(t)=-<\Psi_0| \delta \hat S_z(t)
\hat S_z|\Psi_0> =(1 - |\alpha(t)|^2)/2. 
\label{C_0}
\end{equation}
Then, inserting $\Psi(t)$ into the Schr\"odinger equation we obtain
\begin{eqnarray}
 i\frac{d\alpha(t)}{dt} &=& -\frac{1}{4} A
\alpha(t) + \sum_k \frac{A_k}{2}\beta_k(t)  -\frac{\epsilon_z \alpha(t)}{2},
\nonumber
\\  i\frac{d\beta_l(t)}{dt} &=& (\frac{A}{4}-\frac{A_l}{2} )\beta_l(t)  +
\frac{A_l}{2} \alpha(t)+\frac{\epsilon_z \beta_l(t)}{2},
  \label{system}
\end{eqnarray}
 where $A = \sum_k A_k $. 
Laplace transforming (\ref{system}),
 $\alpha(\tilde\omega)= \int_0^{\infty} dt \exp(-\tilde \omega t) \alpha(t)$, we obtain 
\begin{equation}
\alpha(\tilde\omega) = \frac{i\alpha(t=0^+)}{i\omega + A'/2 - \sum_k A_k^2/4(i\omega + A_k/2)},
\label{alpha}
\end{equation}
where $i \omega = i\tilde \omega -A'/4$,  $A'= A + 2\epsilon_z$, and we have used the fact that $\beta_k(t=0)=0$. 
Now we use the identity $\sum_k A_k^2/4(i\omega + A_k/2)= A/2 - i\omega \sum_k A_k/2(i\omega + A_k/2)$ and  replace the sum 
$\sum_k \frac{A_k}{i\omega + A_k/2}$ over the  $xy$-plane   
 by an integral.  Calculating it  using 
$|\Psi({\bf r}_k)|^2$ given in Eq.(\ref{func}) 
we obtain:
\begin{equation}
\alpha(t)= \frac{1}{2\pi i} \int_{\Gamma} 
\frac{d \omega\, i \exp[(\omega -iA'/4 )t]}
{i\omega+ \epsilon_z + \pi N i\omega  \int
dz\ln (1-
\frac{iA \chi_0^2(z)}{2\pi N \omega})}.
\label{Zeeman}
\end{equation} 
   As usual, the integration contour $\Gamma$
in Eq.(\ref{Zeeman}) is the vertical line in the complex $\omega$-plane so that
all singularities of the integrand  lie to its left (see Fig. 3).  These singularities are: 
two branch points ($\omega=0, \/
\omega_0=iA 
\chi_0^2(0)/2\pi N \equiv iA_0/2$),  and first order poles which lie on the imaginary axis
($\omega=iv$).  For  $\epsilon_z>0$ there is  one pole, while for 
 $\epsilon_z<0$ there  are two poles, and  for $\epsilon_z=0$ there is one 
first order pole at $\omega_1 \approx iA/2 + iA \int dz \chi_0^4(z) /4\pi N$. 
For the contribution from the branch cut between $\omega =0$ and $\omega =
\omega_0$ we obtain 
\begin{eqnarray}
&\tilde \alpha(t)& = \frac{e^{-iA't/4}}{\pi N} 
\int_0^1 {d\kappa  2z_0 \kappa e^{i\tau' \kappa}}  
\Big \{[\kappa \int dz \ln|1 - \frac{\chi_0^2(z)}{\chi_0^2(0) \kappa}| \nonumber \\
&+&  
\kappa/\pi N - 2\epsilon_z/A\chi_0^2(0)]^2 + (2\pi z_0)^2\kappa^2\Big \}^{-1},
\label{arbitr}
\end{eqnarray}  
where $\tau'= \tau \chi_0^2(0)$, 
and  $z_0=z_0(\kappa)$ is defined through
$\chi_0^2(z_0) = \chi_0^2(0) \kappa$. 
We have introduced the dimensionless variable $\kappa = \omega/\omega_0 \leq 1$. In terms of this new variable the argument of the log function in Eq.(\ref{Zeeman}) has the form $1-\chi_0^2(z)/\kappa \chi_0^2(0)$. 
Therefore, for a given $\kappa$ this expression changes the sign at $z_0(\kappa)$ which can be found from the equation given above. 
\par
Thus, the physical picture can be described as follows. 
At the initial time $t=0$ the system has some energy corresponding to the pole and starts to oscillate back and forth each time visiting different frequencies within the branch cut which corresponds to the flip-flop processes with the nuclei located at different sites. Therefore, the contribution from the branch cut describes the electron spin decoherence. At $\tau$ of the order of unity (where the decay mainly occurs) the 
decoherence is due to the interaction with the nuclei located at distances of the order of the dot radius where the derivative of the coupling constant is maximal. 
For longer times, $\tau \gg 1$, the asymptotics is determined either by the interaction with the nuclei located far from the dot or near the dot center depending on the Zeeman field value. 

First we consider the case $\epsilon_z=0$, i.e. no magnetic field.  The asymptotic behavior
 of  Eq.(\ref{arbitr}) for
$\tau \gg 1$ is determined by  $\kappa \ll 1$. For example, for
$\chi_0^2(z)/\chi_0^2(0)= \exp(-z^2)$ we find 
\begin{equation}
\tilde\alpha \propto 1/\ln^{3/2}\tau . 
\end{equation}
  This behaviour is not universal and is determined by the form of the electron wave function at distances that are large compared to the dot size  since at large $\tau$ the decoherence is due to the interaction with the nuclei located far from the dot. As a result, a nuclear spin diffusion induced by the hyperfine interaction occurs in that region. 
\par
Thus, the decay  of $|\alpha(t)|$ starts at $\tau >1$,  i.e. at $t>
N/A$, as in the unpolarized case (see Fig. 4).  Note that the magnitude of
$\tilde \alpha$ is of  order  of $1/N$ (see also Ref.\onlinecite{BLD}), thus, the decaying part of the initial
spin state has this smallness as well, in contrast to 
the unpolarized case above where this part is of order one. 
The reason for this smallness is that for a fully polarized state the gap seen by the electron spin through the hyperfine interaction is $A/2$, therefore, only a small portion $\sim 1/N$ of the opposite (+1/2) state can be admixed. Indeed, in this case the change of the energy of the electron subsystem is $\sim A/N$ which can be compensated by the energy (order of $A/N$) of a magnon excited in the nuclear subsystem.  

 For a large Zeeman field ($|\epsilon_z| \gg
A$) and for $\tau \gg 1$, the main contribution in (\ref{arbitr}) is given for  $\kappa
\rightarrow 1$, i.e.  by the interaction with the nuclei located near the dot center. 
 Expanding
$\chi_0^2(z)$ for small $z$ (see above), we obtain $z_0^2= 2\chi_0^2(0)
(1-\kappa)/(\chi_0^2)''$. Then from Eq.(\ref{arbitr}) we have for
$|\epsilon_z| \gg A$
 \begin{equation}
\tilde \alpha(\tau \gg 1)= \frac{ -e^{i\tau'-iA't/4 }}{4\sqrt{\pi} N} 
 \frac{\chi_0^2(0)}{\sqrt{(\chi_0^2)''}}\frac{A^2}{\epsilon_z^2} 
\frac{(1+i)}{\tau^{3/2}}.
\label{arb}
\end{equation}
 From this we find then that  the
correlator (\ref{C_0})
agrees with the perturbative result (\ref{cor1}) for the fully polarized state,
i.e.  $ C_0(t)-C_0(\infty) \propto 1/t^{3/2}$. 
(Note that the asymptotic
behaviour of the correlator  is given by the term
which is the cross product of the pole contribution and the one (Eq.(\ref{arb})) from the
branch cut.)
This agreement is to be expected, since for a
large Zeeman field, the perturbative treatment with a small parameter
$A/|\epsilon_z| \ll 1$
 is meaningful (the same is true for any model with a small expansion
parameter, for example, for a system with 
anisotropy, where 
the hyperfine constants in perpendicular and transverse
directions are different, see  section \ref{anisotropy}).
However, at zero Zeeman field, when the system cannot be treated
perturbatively, we find  $C_0(t)- C_0(\infty) \propto 1/\ln^{3/2}t$, and the agreement
with (\ref{cor1}) breaks down.  Nevertheless, the characteristic time scale 
for the onset of the non-exponential decay is  the same for all cases and given by
$(A/N)^{-1}$. We have also checked  that  $<\hat S_x(t)-\hat S_x(0)>$ has the same behaviour as $
<\hat S_z(t)-\hat S_z(0)>$ with the same characteristic time - N/A. 

\par 
Finally we mention that the observed decay of the electron spin can be experimentally studied by  local NMR  measurements \cite{experim}.  

\subsection{Fully polarized nuclei. 2D case}
\label{2D}

Here we consider  the 2D case when there is no variation of the coupling constants in one direction, i.e.  
we use the model representation $\chi_0^2(z)= \theta (1/2- |z|)$. 
This case allows us to follow the dependence of the electron spin  decay law on the 
spatial variation of the electron wave function in different directions. 
From Eq.(\ref{Zeeman}) we obtain the following singularities:  two branch points ($\omega=0, \/ \omega_0=iA/2\pi N$) and the first order poles which lie on the imaginary axis ($\omega=iv$). 
The position of these poles can be found from the equation: 
\begin{equation}
\exp(-\frac{1}{\pi N}) \exp(\frac{2\epsilon_z \xi}{A}) = 1-\xi; \/\/ \xi= A/2\pi N v.
\end{equation} 
For positive $\epsilon_z$ there is only one solution of this equation, for negative 
 $\epsilon_z$ - there  are two. 
Thus, at $\epsilon_z=0$ there is one  first order pole at $\omega_1= \frac{iA}{2\pi N} \frac{1}{[1- \exp(-1/\pi N)]}$. 
For the contribution from the branch cut between $\omega =0$ and $\omega = \omega_0$ we obtain: 
\begin{widetext}
\begin{equation}
\tilde \alpha(t)= \frac{\exp(-iA't/4)}{\pi N} 
 \int_0^1 \frac{d\kappa \kappa \exp(i\tau \kappa)}{[\kappa \ln(-1 + 1/\kappa) + 
(\kappa/\pi N) - (2\epsilon_z/A)]^2 + \pi^2\kappa^2}.
\label{Zeemtilde}
\end{equation} 
\end{widetext}
Let us consider first the case of  zero Zeeman field. Then we have $\alpha(t>0)= \exp(-iAt/4)\exp(\omega_1 t) /(\pi N [\exp(1/\pi N) -1])  + \tilde \alpha (t)$. 
 The asymptotic behaviour of Eq.(\ref{Zeemtilde}) at $\epsilon_z=0$ and for $\tau \gg 1$ is $1/\ln\tau $. 
 In the case of a strong Zeeman field, $|\epsilon_z| \gg A$, 
 it follows from Eq.(\ref{Zeemtilde}) that the asymptotics at $\tau \gg 1$ is $1/\tau$. 
Note, that this asymptotics is true  in the interval $1\ll \tau \ll \exp(2|\epsilon_z|/A)$.
Then we obtain:  
\begin{widetext}
\begin{equation}
<\hat S_z(t)> = \frac{1}{2} - |\alpha|^2 =
-\frac{1}{2}[1-\frac{A^2}{2\pi N \epsilon_z^2}] - \frac{A^2}{2\pi N \epsilon_z^2} Re\{\frac{\exp(-iA't/2) \exp(i\tau)}{i\tau}  \}.
\label{S}
\end{equation}
\end{widetext}
The change of the asymptotics occurs even at $|\epsilon_z| \ll A$. If $1\ll \tau/\ln \tau \ll A/|\epsilon_z|$, then the asymptotics is as before, i.e.  $1/\ln \tau$.
For $\tau/\ln \tau \gg A/|\epsilon_z| \gg 1$ it is  $1/(\tau \ln^2\tau)$. 

\subsection {Anisotropy in the exactly solvable model.}
\label{anisotropy}

We consider here the model where the hyperfine constants in the $z$-direction and in the transverse direction are different (there is no Zeeman field): 
 $A_i^z =A_z v_0 \mid \Psi({\bf r}_i) \mid ^2$, 
$A_i^{\perp} =A_{\perp} v_0 \mid \Psi({\bf r}_i) \mid ^2$.
Then from a system of the equations similar to Eq.(\ref{system}) we obtain 
the following solution (again for the 2D case, when $\chi_0^2(z)= \theta (1/2- |z|)$):
 \begin{widetext}
\begin{equation}
\alpha(t)= \frac{i}{2\pi i} \int_{\Gamma} 
 d\omega \frac{\exp(-iA_zt/4) \exp(\omega t) }{i\omega+ (A_z/2) - (A_{\perp}^2/2A_z) + \pi N (A_{\perp}^2/A_z^2) i\omega  \ln (1- \frac{iA_z}{2\pi N \omega})}.
\label{Anisot}
\end{equation}
\end{widetext}
Note that this equation has the same form as Eq.(\ref{Zeeman}), i.e. an anisotropy plays the role of a Zeeman field. Then for the contribution from the branch cut (the decaying part of the initial spin state) we have:
\begin{widetext}
\begin{equation}
\tilde \alpha(t)= \frac{\exp(-iA_zt/4)}{\pi N \sigma} 
 \int_0^1 \frac{d\kappa \kappa \exp(i\tau \kappa)}{[\kappa \ln(-1 + 1/\kappa) + 
(\kappa/\pi N \sigma) +1 - 1/\sigma]^2 + \pi^2\kappa^2},
\label{Anistilde}
\end{equation} 
\end{widetext}
where $\sigma= A_{\perp}^2/A_z^2$, and  $\tau$ is defined now as 
$\tau= A_zt/2\pi N$. In the case of strong anisotropy, $\sigma \ll 1$, we  have 
at $\tau \gg 1$:
\begin{widetext}
\begin{equation} 
<\hat S_z(t)>  = \frac{1}{2} - |\alpha|^2 =
-\frac{1}{2}(1-\frac{2\sigma}{\pi N}) - \frac{2\sigma}{\pi N} Re\{\frac{\exp(-iA_zt/2) 
\exp(i\tau)}{i\tau} \}.
\end{equation} 
\end{widetext}
The same result follows for the polarized state from Eq.(\ref{4}) in  the perturbative approach.

\subsection{Some features of the fully polarized state.}
\label{features}

There are several interesting features which we can observe for the fully
polarized state. In an external Zeeman field, the effective gap seen by the
electron spin  is $A'/2= A/2 + \epsilon_z$. Thus,  when
$\epsilon_z$ is made negative this gap decreases and even vanishes
at $ |\epsilon_z|= A/2$. From Eq.(\ref{Zeeman}) we find that 
the two poles are symmetric in
this case, and the system resonates between the two frequencies $\omega_{\pm} = \pm i
A(\int  \chi_0^4(z)dz)^{1/2}/\sqrt{8\pi N}$.  Note that the 
residual gap is of  order  $A/\sqrt{N}$ (and not $A/N$, as one might
naively expect).    Near this Zeeman field we have
$|\alpha(t)|^2=\cos^2(\omega_{+}t)$ (up to small corrections of order $1/N$),
and, as a result, $|\alpha|^2$ averaged over time  is 1/2, i.e. the
up and down states of the electron spin are strongly coupled via the nuclei (see Fig. 5). In contrast,
outside this resonance regime  the value  of $|\alpha|^2$ is
close to 1 (again with small $1/N$ corrections), i.e. 
$<\hat S_z(t)>=1/2-|\alpha|^2$ is close to
-1/2 at any time. The width of the resonance is $\sim A/\sqrt{N}$, i.e. small compared to
the initial gap $A/2$.    We note that this behavior represents
periodic (Rabi) oscillations with a single well-defined frequency and
is not related  to decoherence. [The latter is described by the branch cut 
contribution $\tilde \alpha$ which  remains small (order 1/N) even near 
the resonance.] This abrupt change  in the 
amplitude of oscillations of $<\hat S_z(t)>$ (when changing $\epsilon_z$ in
a narrow interval around $A/2$) can be used for an experimental detection of the fully
polarized state. Note that the weight of the upper pole alone (i.e. that which exists at
$\epsilon_z=0$) also  drops abruptly from a value close to 1 to a value much smaller
than 1 in the same narrow interval, which can be
experimentally checked by Fourier analysis.  
\par
Another special value of the Zeeman
field corresponds to the case when the  upper pole is close
to $\omega_0$  ($\kappa
=1$)-- the upper  edge of the branch cut.  This occurs (see Eq.(\ref{Zeeman})) at
the  critical value
$\epsilon_z^{\star}=bA/2<0$, where $b= \chi_0^2(0)\int dz \ln
|1-\chi_0^2(z)/\chi_0^2(0)| < -1$ is a non-universal number which depends on the
dot shape.    Since at a finite Zeeman field the asymptotics in $t$ is
determined by $\kappa$'s close to 1,  we see from Eq.(\ref{arbitr}) that
for $\epsilon \approx \epsilon_z^{\star}$ the asymptotics  changes
abruptly. Indeed, for $((\epsilon_z-\epsilon_z^{\star})/A)^2 \ll 1$,  we find
$\tilde\alpha \propto 1/\sqrt{\tau}$, for 
$ 1\ll \tau \ll  ((\epsilon_z-\epsilon_z^{\star})/A)^{-2}$, and $\tilde\alpha
\propto
1/\tau^{3/2}$, for
$\tau \gg ((\epsilon_z-\epsilon_z^{\star})/A)^{-2}$. Thus,  when approaching the
critical Zeeman field $\epsilon_z^{\star}$ there is a {\em slow down} 
of the asymptotics
from $1/\tau^{3/2}$ to
$1/\tau^{1/2}$.  It is interesting that this slow down is related to a strong
modulation of the density of states (DOS) of the excitations within the continuum band
(branch cut) near its edge when $\epsilon_z \rightarrow \epsilon_z^{\star}$. 
In the subspace of none or one
 nuclear spin flipped (see
Eq.(\ref{function})), the DOS $\nu$ becomes  
\begin{equation}
\nu(u)= \frac{1}{\pi}Im [G_0(u) + \frac{d}{du} \ln D(u)],
\label{DOS}
\end{equation}
 where $u=i\omega$, $G_0(u)= \sum_k 1/(u + A_k/2)$ is the "unperturbed
Green's function", and $D(u)$ is the denominator of $\alpha(\omega)$, see Eq.(\ref{alpha}).
\cite{Kosevich} The derivation of Eq.(\ref{DOS}) is given in the Appendix. 
One can then show that for $\epsilon_z \rightarrow \epsilon_z^{\star}$
(i.e. the upper pole
approaches the continuum  edge), the DOS  develops a square
root singularity: $\nu(u) \propto 1/\sqrt{\omega_0 - u}$. Simultaneously,  the weight
of the upper pole vanishes linearly in $\epsilon_z$ as
$\epsilon_z^{\star} -\epsilon_z\rightarrow 0$.  
 
Finally, the nuclear state is characterized by $\beta_k(t)$, which allows for similar
evaluation as for $\alpha$. Here we just note that its branch cut part,
$\tilde \beta_k(t)$, is nonmonotonic 
in time, particularly pronounced at $\epsilon_z \rightarrow \epsilon_z^{\star}$: 
First, $\tilde \beta_k(t)$ grows like $ \sqrt\tau $, until $\tau$
reaches
$\sim 1/(1-a_k)\gg 1$, and
then  it decays like $ 1/(\sqrt{\tau}(1-a_k))$, with $a_k=A_k/A_0
\rightarrow 1$. Thus,  $\beta_k$ is  maximal  for  $A_k$
close to $A_0$, i.e. the nuclei near the dot-center are 
affected most by the hyperfine
interaction with the electron spin.

\section {Dependence of the electron spin decoherence on the initial nuclear state.} 

So far we have assumed that the initial nuclear state 
had the form of a single tensor product state. 
Now we study the dependence of the electron spin decoherence on the initial nuclear state and start with the simple case of homogeneous coupling. 

\subsection{Homogeneous coupling.}
\label{homo}

In this subsection we consider the case of homogeneous coupling, when all the coupling constants are equal, i.e. $A_k = A/N$.  It is instructive to start with it, since in contrast to naive expectations even in this simple case there is some time dependence of $<S_z(t)>$ which cannot be described by a single frequency. On the other hand, in this case the expectation value $<S_z(t)>$ is a periodic function and, for that reason, the electron spin dynamics is coherent though the corresponding period can be very large for large $N$ (see below).  The Hamiltonian is very simple, $ \hat H =(A/N) {\bf I}\cdot {\bf S}$, where 
${\bf I}$ is the total nuclear spin. Then, we can diagonalize $\hat H$ and obtain the result:
\begin{eqnarray}
E_1 = \frac{AI}{2N} \,\,\, , for J_1 = I +1/2, \nonumber \\
 E_2= -\frac{A(I+1)}{2N}\,\,\, , for J_2=I-1/2.
\label{energy}
\end{eqnarray}
In the simplest case of a fully polarized nuclear state and opposite initial electron spin polarization (with   
 $\Psi_0$ given in Eq.(\ref{0})) we obtain the following result:
\begin{equation}
<S_z(t)> = -\frac{1}{2} + \frac{2N}{(N+1)^2} (1-\cos \Omega_{N/2}t),
\label{N/2}
\end{equation}
where $\Omega_{N/2}=A(1+1/N)/2$. This result can be easily understood, since for the homogeneous coupling 
$I^2$ is conserved and the initial nuclear state corresponds to the maximal value $I=N/2$. Then from Eq.(\ref{energy}) we obtain for the difference of $E_1$ and $E_2$ corresponding to $I=N/2$ the  value which is equal to $\Omega_{N/2}$.  
Note that the magnitude of the oscillating term is $1/N$ for $N\gg 1$, as it has been already observed in Sec. \ref{polarized}.

In the case when initially one nuclear spin is flipped, the exact ket
has the form
 \begin{eqnarray}
\Psi_1(t)= \sum_k \alpha_k(t) |\Downarrow; \uparrow, \uparrow,
\downarrow_k,\uparrow,...> + \nonumber \\
+ \sum_{k>l}d_{kl}(t)|\Uparrow; \uparrow,\uparrow,\downarrow_k,\uparrow,\downarrow_l,...> ,
\label{Psi_1}
\end{eqnarray}
and the normalization condition is
$\sum_k |\alpha_k(t)|^2 + 
\sum_{k>l}|d_{kl}(t)|^2 =1$, with initial condition $d_{kl}(0)=0$.  
From the solution of the Schr\"odinger equation we obtain  for $<S_z(t)>$
\begin{widetext}
\begin{equation}
<S_z(t)> = -\frac{1}{2} +\frac{2(N-2)}{(N-1)^2}(1-\cos \Omega_{N/2-1}t)- 
\frac{2|\alpha(0)|^2}{N} \left [\frac{(N-2)}{(N-1)^2}(1-\cos \Omega_{N/2-1}t)-\frac{2(N-1)}{(N+1)^2}(1-\cos \Omega_{N/2}t)\right ],
\label{N/2-1}
\end{equation}
\end{widetext}
where  $\Omega_{N/2-1}=A(1-1/N)/2$ is the frequency corresponding to the solution $E_1-E_2$ of Eq.(\ref{energy}) with $I=N/2-1$, and $\alpha(0)= \sum_k \alpha_k(0)$. Now, in contrast to the fully polarized case there are oscillating terms with two frequencies involved. This is because the initial nuclear state corresponds to $I_z =N/2-1$. This value of $I_z$ can be realized with $I=N/2$ and $I=N/2-1$. Note that  
for $N\gg 1$ the amplitude of the oscillating term corresponding to $I=N/2$ is  smaller by factor $1/N$ than the 
amplitude of the oscillating term corresponding to $I=N/2-1$. This is a general rule which can be checked for an arbitrary initial state with $I_z=n$.  This state is constructed mostly from the state with $I=n$ and the contributions of all other $I$ values are small for $N \gg 1$.  The expectation value 
$<S_z(t)>$, Eq.(\ref{N/2-1}), is periodic with  period $T=4\pi N/A$. 
The initial condition enters Eq.(\ref{N/2-1}) through the quantity 
 $\alpha(0)= \sum_k \alpha_k(0)$. We can easily check that in the  case of the homogeneous coupling there is only  a  weak dependence on the type of the initial nuclear state. 
For example, the cases $\alpha_k(0)=\delta_{kn}, \alpha(0)=1$ (which is a single tensor product state) and
a randomly correlated (entangled) state when the coefficients $\alpha_k(0)$
have random phases, so that $|\alpha(0)| \ll \sqrt{N}$, correspond to the same solution $<S_z(t)>$ 
for $N\gg 1$, as is easily seen from Eq.(\ref{N/2-1}).  
\par
The solution for the initial nuclear state with a larger number of flipped spins can be obtained similarly.
For example, when the initial state corresponds to two flipped nuclear spins, the solution for $<S_z(t)>$ contains  oscillating terms with three frequencies, $\Omega_{N/2}, \Omega_{N/2-1}, \Omega_{N/2-2}$. 
These frequencies are the solutions of Eq.(\ref{energy}) for $I=N/2, N/2-1, N/2-2$. 
Moreover, the initial conditions (the information about the initial nuclear state) enter now 
$<S_z(t)>$
 in a more complicated way besides the quantity $|\alpha(0)|^2$ which we observed for the case with one flipped spin, since it 
 contains also the quantity $\sum_k |\alpha_k(0)|^2$. 
Again, depending on the type of the correlations between the coefficients $\alpha_k(0)$ we can have different dynamics of the electron spin. However, the dependence on the type of the initial nuclear state is again rather weak.

\subsection{Inhomogeneous coupling.}
\label{initial}

 Up to now we have only considered the decoherence of the electron spin caused by inhomogeneous coupling for a given initial nuclear state which 
had the form of a single tensor product state. 
Here we consider a more general initial nuclear state which can be entangled, i.e. contains the {\it coherent} superposition of the single tensor product states 
$\sum_{T}\alpha_{T}|T\rangle$, where the sum goes over
the tensor product states $|T\rangle$. 
 This problem was addressed numerically in Ref.\onlinecite{Schliemann}.  We give here the exact analytical solution of a typical problem of this kind which consists of a very simple initial nuclear state  but still contains  the relevant physics. In particular, we can examine the dependence of the electron spin decoherence (and the corresponding time scales) on the type of 
initial  state.  As an example of the initial electron and nuclear state which we can treat exactly we choose the following one:
All but one nucleus are polarized in  z-direction and the 
electron spin 
 is in the up state. 
We will see that the results are the same for all cases, namely, for this particular nuclear state the time scale for the onset of the decay is always $\sim N/A$ independent  of the phases
of the coefficients $\alpha_{T}$.

The {\it exact}  ket of the system for $t>0$ has the form
\begin{equation}
\Psi(t)= \sum_k \alpha_k(t) |\Uparrow; \uparrow, \uparrow,
\downarrow_k,\uparrow,...> + \beta(t)|\Downarrow; \uparrow,\uparrow,\uparrow,...> ,
\label{function1}
\end{equation}
  where the first term at $t=0$ represents the initial state of the system. The normalization condition is $\sum_k |\alpha_k(t)|^2 + 
|\beta(t)|^2 =1$, and  $\beta(t=0) =0$.
The dependence on the initial nuclear state comes from a different realization of the coefficients
$\alpha_k(0)$. If, for example, we have $\alpha_k(0)=\delta_{kn}$, then  the initial nuclear state consists of a single product state. On the other hand, the entanglement of the initial nuclear state can be of a different kind, depending on whether  
the coefficients $\alpha_k(0)$ have random phases  or correlated phases.  
From the system of equations similar to Eq.(\ref{system}) we obtain: 
\begin{equation}
\beta(t)=  e^{-itA/4}\int_{\Gamma} \frac{d \omega}{2\pi i}
 \exp(\omega t) \beta(\omega), 
\end{equation}

\begin{equation}
\beta(\omega)= \frac{i}{2}
\frac{1}{i\omega + A/2 -\sum_i A_i^2/4(i\omega +A_i/2)}\sum_k \frac{A_k \alpha_k(0)}{i\omega + A_k/2}. 
\label{beta}
\end{equation} 
Here again $A=\sum_k A_k$ and the  integration contour $\Gamma$ 
 is the same as in Eq.(\ref{Zeeman}), see Fig. 3.  
The quantity $<\hat S_z(t)>$ we are interested in is 
given by the equation 
\begin{equation} 
<\hat S_z(t)>  = \frac{1}{2} - |\beta(t)|^2. 
\label{S_z}
\end{equation} 
As it was already mentioned above,  Eq.(\ref{beta}) contains  the information about the initial nuclear state through the coefficients $\alpha_k(0)$. Note that the denominator of the first factor in the solution for $\beta(\omega)$,
 see Eq.(\ref{beta}), is exactly the same as that in  $\alpha(\omega)$, Eq.(\ref{alpha}), 
if we put in the latter $\epsilon_z=0$.  Thus,   Eq.(\ref{beta}) contains  partially the same singularities as 
 $\alpha(\omega)$. There are, however, some additional singularities whose character depends on the properties of the coefficients $\alpha_k(0)$.

\par
1. Let us consider the case $\alpha_k(0)=\delta_{kn}$. This means that the initial nuclear state is just a simple tensor product state. Then besides the branch cut and the first order pole (outside the branch cut) which we had before for $\alpha(\omega)$, there is an additional first order pole $\omega=iA_n/2$ which lies within  the branch cut. 
Considering again the continuous limit (i.e. replacing the sum by an integral and using Eq.(\ref{func})) we obtain:
\begin{widetext}
\begin{equation}
\beta_n(t)= \frac{e^{-itA/4}}{4\pi i} \int_{\Gamma} 
\frac{d \omega}{\omega}  e^{\omega t} 
\frac{A_n}{(i\omega + A_n/2)}
\frac{1}{[1 + \pi N  \int
dz\ln (1-
\frac{iA \chi_0^2(z)}{2\pi N \omega})]}\/.
\label{beta_n}
\end{equation}
\end{widetext}
We have evaluated  Eq.(\ref{beta_n}) in the 2D case with the model function $\chi_0^2(z)$
 introduced in Sec.
\ref{2D} and find
\begin{widetext}
\begin{equation}
\beta_n(\tau)= e^{-itA/4}\frac{x_0}{\pi N}\left [P \int_0^1 \frac{dx e^{ix\tau}}{x (x_0-x)[\pi^2 + \ln^2(1/x-1)]} - e^{i\pi N \tau + i\tau/2} - \frac{\ln (1/x_0 -1)e^{ix_0 \tau}}{x_0[\pi^2 + \ln^2(1/x_0 -1)]} \right ]\/, 
\label{tildebeta}
\end{equation}
\end{widetext}
where $x_0=A_n/A_0 <1$, and $P$ means the principle value of the integral. Since $\beta_n(\tau)$ is
of order  $1/N$, it follows from Eq.(\ref{S_z}) that the decaying part of the electron spin state has now a smallness $\sim 1/N^2$. This is simply due to the fact that initially the electron has the same spin orientation as all the nuclei except one, and 
the flip-flop process is only allowed  with this particular nucleus.  However,
the time scale for the onset of the decoherence is the same as before, i.e. it starts at $\tau >1$ and the asymptotic dependence of Eq.(\ref{S_z}) at $\tau \gg 1$ is $1/\ln \tau$, the same as before, see Sec. \ref{2D}.  

\par
2. Let us now consider the case when $\alpha_k(0)=1/\sqrt{N}$. This corresponds to the  entangled initial nuclear state where all the terms have the same phases. It is easy to see that the singularity of the second factor in Eq.(\ref{beta}) (a branch cut) coincides with that of the first factor. 
Thus, we obtain that the decay starts at the same time $\tau >1$ as it was for the single tensor product state
with the same decay law, i.e. $1/\ln\tau$. 

\par
3. Finally, we consider the case when the phases of the coefficients $\alpha_k(0)$ are random.  
It is  obvious that while the decay law can be different depending on the particular choice of the 
coefficients $\alpha_k(0)$ (phases), the characteristic time scale for the onset of the decay is always the same, i.e. $N/A$. This follows from the fact that the singularity of the first factor in Eq.(\ref{beta}) (a branch cut) is exactly the same as of the second factor. 
This in turn is the consequence of the fact that in the model considered here it is only one nuclear spin which is flipped. Therefore, the characteristic energy change which determines the time scale for the onset of the decay is $A_k \simeq A/N$. 
It is also clear that in a more general case, for example, of an unpolarized nuclear state there  is a different energy scale involved which is related to different values of the magnetic fields corresponding to different configurations which are present in the initial superposition of the states. This scale presumably is  $A/\sqrt{N}$. 
Then the result can be entirely different. It has actually been observed in Ref.\onlinecite{Schliemann} that there is a strong dependence of the characteristic time scale for the onset of the electron spin decay on the type of the initial nuclear spin state.

 \section {Averaging over nuclear configurations. Dephasing time for an ensemble of dots.} 
\label{ensemble}

 In Secs. \ref{unpolarized} and \ref{polarized} we have seen  that the  decay
of  $C_n(t)$ occurs in the time interval $ N/A \ll t \ll
N^2/A$, with $N/A \simeq 10^{-6} s$ in GaAs dots. On the other hand,
the electron spin precesses in the net nuclear field (see Eq. (\ref{cor1}))
with the
characteristic period 
$(h_z)_n^{-1} \simeq \omega_N^{-1} 
\simeq 10^{-8} \div 10^{-9} s$.
Thus, $\omega_N^{-1}\ll N/A$, and we see that
the electron spin undergoes many  precessions   in a given
nuclear field before decoherence sets in due to the non-uniform hyperfine couplings
$A_k$.  
This  behavior changes dramatically when we average over  nuclear
configurations \cite{Efros,ensemble}.
For that purpose we consider high temperatures, $k_BT \gg \hbar \omega_N$,
and average $C_n(t)$ in Eq.(\ref{4}) over all
 nuclear configurations, i.e. $C(t)=\sum_n C_n(t)/\sum_n$. We then find
\begin{equation} 
C(t)=
\sum_k \frac{-A_k^2}{8} \int_0^t dt_1\int_0^t   dt_2
\prod_{i\neq k}\cos[\frac{A_i}{2}(t_1-t_2)]. 
\label{aver}
\end{equation} 
For $\tau \ll 1$, we get $\prod_{i\neq
k}\cos(A_it/2)= \exp[- NC (At/2\pi N)^2 ]$, where 
$C= \pi\int dz \chi_0^4(z)/4$.  Thus, the averaged spin correlator $C(t)$,
Eq.(\ref{aver}), is of  order 
$- \int_0^{\omega_N t} dx \Phi (x)$, with 
$\Phi$ being the error function.  Thus, $C(t)$ 
grows without bound as $ \omega_N t$
for $ \omega_N t \gg 1$ (the condition $\tau \ll 1$ can still be 
satisfied). Consequently, the perturbative approach breaks  down  even
in leading order in $\hat V$ (we recall that {\em without} averaging the 
divergences  occur in all higher but not in lowest order).
To treat this case properly, we need a non-perturbative approach.
For that purpose we  calculate now the correlator $C(t)$
exactly by treating the nuclear field  purely classically, i.e. as a c-number.  Then we
obtain,  
\begin{equation}
 C_n(t) = - \frac{h_{N\perp}^2}{4 h_N^2} (1- \cos h_N t), 
\label{classic}
\end{equation} 
where $h_N = \sqrt{h_{Nz}^2 + h_{N\perp}^2}$ is the nuclear field, with $ \/
h_{N\perp}^2= h_{Nx}^2 + h_{Ny}^2$. The value of  $h_N$  corresponds to a given
nuclear configuration $n$. To make contact with the perturbation procedure  we used
before in the quantum case we go to the regime $h_{N\perp}^2 \ll
h_{Nz}^2$,  where $h_N$ can be replaced by  $h_{Nz}$ in Eq.(\ref{classic}).
Then we average the resulting expression  $(h_{N\perp}^2/ h_{Nz}^2) (1- \cos
h_{Nz} t)$  over a Gaussian distribution for $h_N$, i.e. over $P(h_N) \propto
\exp(- 3h_N^2/2\omega_N^2)$.  The result becomes  proportional to
$\int_{0}^{+\infty} dz \exp(-z^2/2) (1- \cos(\gamma z))/z^2 \propto
\int_0^{\gamma} dx
\Phi(x)$, where $\gamma= \omega_N t/\sqrt{3} $. Thus, we see that we obtain exactly the
same functional form as before from Eq.(\ref{aver}) with the same
divergencies in t. This reassures us that the  treatment of the nuclear
field as a classical field is not essential. On the other hand, the same Gaussian
averaging procedure can now be applied  to the non-perturbative form
Eq.(\ref{classic}). Defining $C_{cl}(t)=\int dh_N P(h_N) C_n(t)$, we obtain 
\begin{equation}
C_{cl}(t)= -\frac{1}{6} [1+ ( \frac{\omega_N^2
t^2}{3} -1)\, e^{-\omega_N^2 t^2/6}].
\label{exact}
\end{equation} 
Thus we get rapid (Gaussian)
 decay  of the correlator for $t\gg \omega_N^{-1}$, giving the dephasing time
$\omega_N^{-1}=\sqrt{N}/A$.
This means that $<\hat S_z(t) S_z>$ saturates at  1/3 of its
initial value of $1/4$. Finally, 
it seems likely that for the case of nuclear quantum  spins 
a non-perturbative treatment of the averaged  correlator $C(t)$
 will lead to a similar rapid time decay as found for the classical case in Eq.
(\ref{exact}).

\section{Conclusion.}

In conclusion, we have studied  the  spin decoherence of an electron
confined to a single
quantum dot in the presence of hyperfine interaction with  nuclear spins. 
  The decoherence is due to a non-uniform coupling of the electron spin to 
nuclei located at different sites. The decoherence time is given by 
$\hbar N/A$ and is of the order of several $\mu s$ for typical GaAs dots. It is shown that in a weak
external Zeeman field the perturbative treatment of the electron spin
decoherence is impossible, in particular, we cannot use the usual  formulas  Eq.(\ref{usual}).  
 Moreover, the decay of the electron spin
correlator in time does not have an exponential character, instead it is given by a power 
or inverse logarithm law.   In the case of a strong Zeeman field the decay has a universal character $\propto 1/t^{d/2}$, where $d$ is the real space dimensionality of the problem. 
We have also solved exactly several model problems which allowed us to investigate the dependence of the 
electron spin decay on the initial state of the nuclear system.  
We have shown that there is a strong  difference between
the  decoherence time for a single dot, 
$\hbar N/A$, and the dephasing time for an ensemble of dots, $\hbar\sqrt{N}/A$.

\section {Acknowledgements.}

We acknowledge  support from the NCCR Nanoscience, 
Swiss NSF, US DARPA and ARO. Part of this work
was performed at the Aspen Center of Physics and at the ITP, UC Santa
Barbara. L.G. acknowledges support from NSF Grant No. DMR-9731756. We thank V. Golovach, J. Schliemann, Al.L. Efros, H. Gassmann, and  F. Marquardt for useful discussions.

\section {appendix}

In this Appendix we derive the 
density of states (DOS) given in Eq.(\ref{DOS}).
The definition for the DOS is 
\begin{equation}
\nu(\tilde \omega)= \frac{1}{\pi} Im Tr \frac{1}{i\tilde\omega -\hat {\cal H}}, \,\,\, i\tilde\omega= E-i\epsilon, \,\,\, \epsilon \rightarrow 0.
\label{DOS1}
\end{equation}
We write the trace as follows: 
\begin{equation}
Tr \frac{1}{i\tilde\omega -\hat {\cal H}}= <\Psi_0|\frac{1}{i\tilde\omega -\hat {\cal H}}|\Psi_0> + 
\sum_k <\Psi_k|\frac{1}{i\tilde\omega -\hat {\cal H}}|\Psi_k>, 
 \label{trace}
\end{equation}
where 
\begin{eqnarray}
|\Psi_0> = |\Downarrow; \uparrow, \uparrow, \uparrow,...>, \nonumber \\  
|\Psi_k> = |\Uparrow; \uparrow, \uparrow,
\downarrow_k,\uparrow,...>, \,\, k=1,..., N.
\label{psi}
\end{eqnarray}
The Laplace transform of Eq.(\ref{system}), $\alpha(\tilde\omega)= \int_0^{\infty} dt \exp(-\tilde \omega t) \alpha(t)$, 
 gives for general initial conditions:
  \begin{equation}
\beta_l(\tilde\omega) = \frac{A_l \alpha(\tilde\omega)/2+ i \beta_l(t=0)}{i\tilde\omega -A'/4 + A_l/2}
\label{beta_l}
\end{equation}
and 
\begin{equation}
\alpha(\tilde\omega) = \frac{i \alpha(t=0)}{D(\tilde\omega)} + \frac{i}{2D(\tilde\omega)}\sum_k 
\frac{A_k \beta_k(t=0)}{i\tilde\omega -A'/4 + A_k/2},
\label{alpha1}
\end{equation}
where 
\begin{eqnarray}
D(\tilde\omega) = i\tilde\omega + A'/4 -\frac{1}{4} 
\sum_k 
\frac{A_k^2}{i\tilde\omega -A'/4 + A_k/2}= \nonumber \\
= i\omega +\epsilon_z + i\omega \sum_k 
\frac{A_k/2}{i\omega + A_k/2}, 
\label{D}
\end{eqnarray}
and $i\omega = i\tilde \omega -A'/4$. 
\par
1) Let us start with the case 
\begin{equation}
\exp(-i\hat {\cal H}t)|\Psi_0> = \alpha(t) |\Psi_0> + \sum_k\beta_k(t) |\Psi_k>, 
\label{c1}
\end{equation}
which corresponds to the initial conditions $\alpha(t=0)=1, \beta_k(t=0)=0$. 
Performing the Laplace transform of Eq.(\ref{c1}), we obtain
 \begin{equation}
<\Psi_0|\frac{1}{i\tilde\omega -\hat {\cal H}}|\Psi_0> =- i\alpha(\tilde\omega)= 1/D(\tilde\omega),
\label{c11}
 \end{equation}
where we have used Eq.(\ref{alpha1}) with initial conditions $\alpha(t=0)=1$, and $\beta_k(t=0)=0$. 
\par
2) Next we consider the case
\begin{equation}
\exp(-i\hat {\cal H}t)|\Psi_k> = \alpha^k(t) |\Psi_0> + \sum_{k'}\beta_{k'}^k(t) |\Psi_{k'}>, 
\label{c2}
\end{equation}
which corresponds to the initial conditions $\alpha^k(t=0)=0$, and $\beta_{k'}^k(t=0)=\delta_{kk'}$. 
Again, performing the Laplace transform of Eq.(\ref{c2}), we obtain
\begin{equation}
\sum_k <\Psi_k|\frac{1}{i\tilde\omega -\hat {\cal H}}|\Psi_k>
=- i\sum_k  \beta_k^k(\tilde\omega).
\label{c21}
 \end{equation}
It follows from Eqs.(\ref{beta_l}, \ref{alpha1}) and the initial conditions indicated above  
that
\begin{equation}
 \beta_k^k(\tilde\omega)=\frac{i}{4D}\frac{A_k^2}{(i\tilde\omega -A'/4 +A_k/2)^2} +\frac{i}{i\tilde\omega -A'/4 +A_k/2}.
\label{beta_kk}
 \end{equation}
Then from Eqs.(\ref{c21}), (\ref{beta_kk}) we obtain
 \begin{widetext}
\begin{equation}
\sum_k <\Psi_k|\frac{1}{i\tilde\omega -\hat {\cal H}}|\Psi_k>= 
\frac{1}{4D}\sum_k \frac{A_k^2}{(i\tilde\omega -A'/4 +A_k/2)^2} +\sum_k \frac{1}{i\tilde\omega -A'/4 +A_k/2}.
\label{c22}
\end{equation}
\end{widetext}
Collecting all the terms, Eqs.(\ref{c11}),(\ref{c22}), we obtain
 \begin{equation}
\nu = \frac{1}{\pi} Im \left[ \frac{1}{D(\tilde\omega )} \left(1+ \sum_k \frac{(A_k/2)^2}{(u + A_k/2)^2} \right) + 
\sum_k \frac{1}{u+ A_k/2}\right ], 
\label{nu}
\end{equation}
where $u=i\omega$. 
From Eq.(\ref{D}) we can easily check that 
 \begin{equation}
\frac{d}{du}\ln D =  \frac{1}{D} \left(1+ \sum_k \frac{(A_k/2)^2}{(u + A_k/2)^2} \right). 
\end{equation}
Thus, from Eq.(\ref{nu}) we finally obtain Eq.(\ref{DOS}) of the main text.


\begin{figure}[h]
\begin{center}
\leavevmode
\includegraphics[width=13cm]{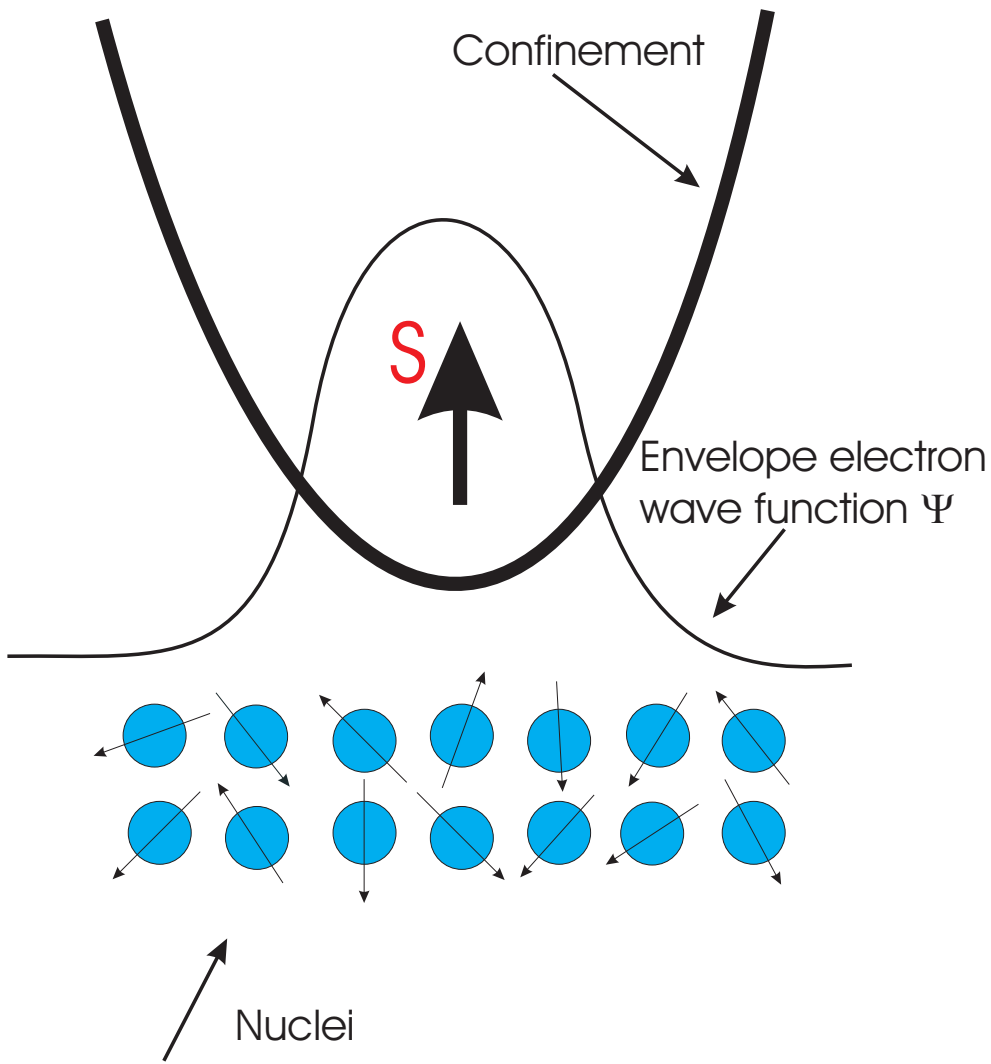}
\end{center}
\caption{A single electron spin localized in a quantum dot described by a parabolic confinement potential (x,y plane). The electron is assumed to be in the orbital ground state described by the envelope wave function $\Psi$, and interacts with the nuclear spins (located at ${\vec r}_i$) via hyperfine interaction $A_i \sim |\Psi({\vec r}_i)|^2$ which varies as function of position ${\vec r}_i$.}
\label{fig:1}
\end{figure}

\begin{figure}[h]
\begin{center}
\leavevmode
\includegraphics[width=13cm]{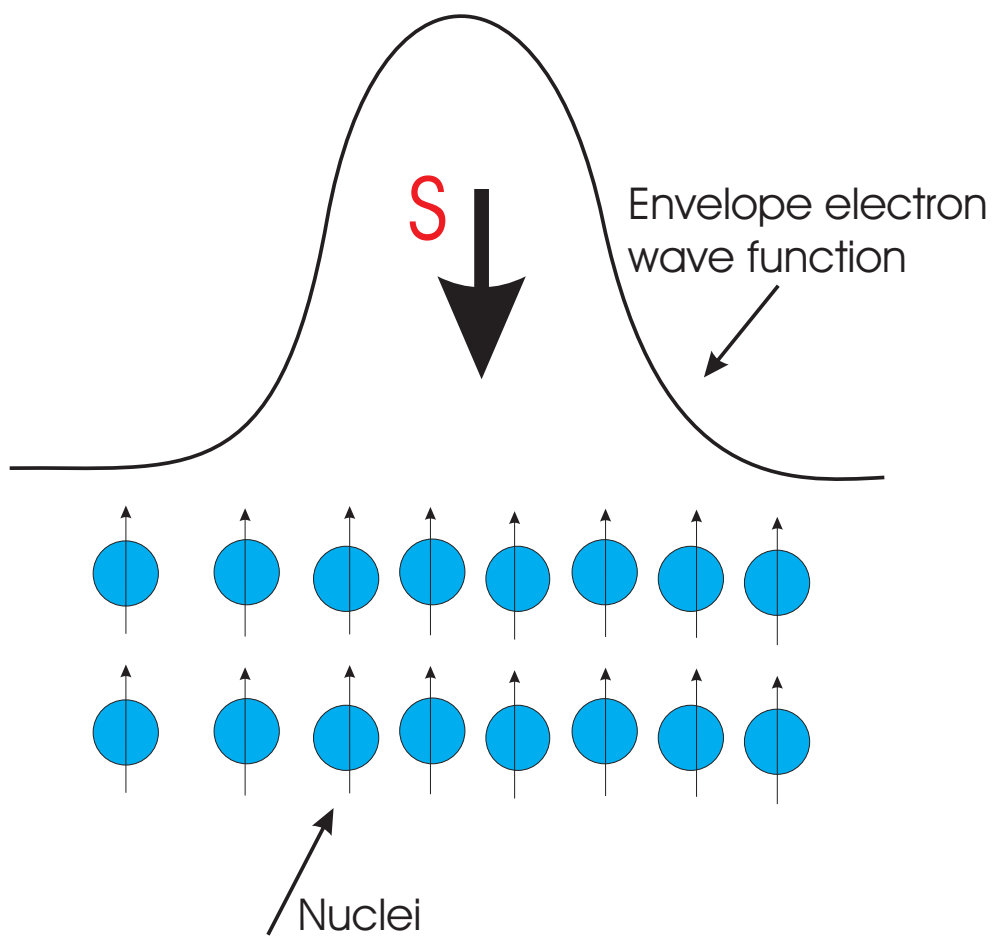}
\end{center}
\caption{The same situation as in Fig.1 but with fully polarized nuclear spins.}
\label{fig:2}
\end{figure}

\begin{figure}[h]
\begin{center}
\leavevmode
\includegraphics[width=13cm]{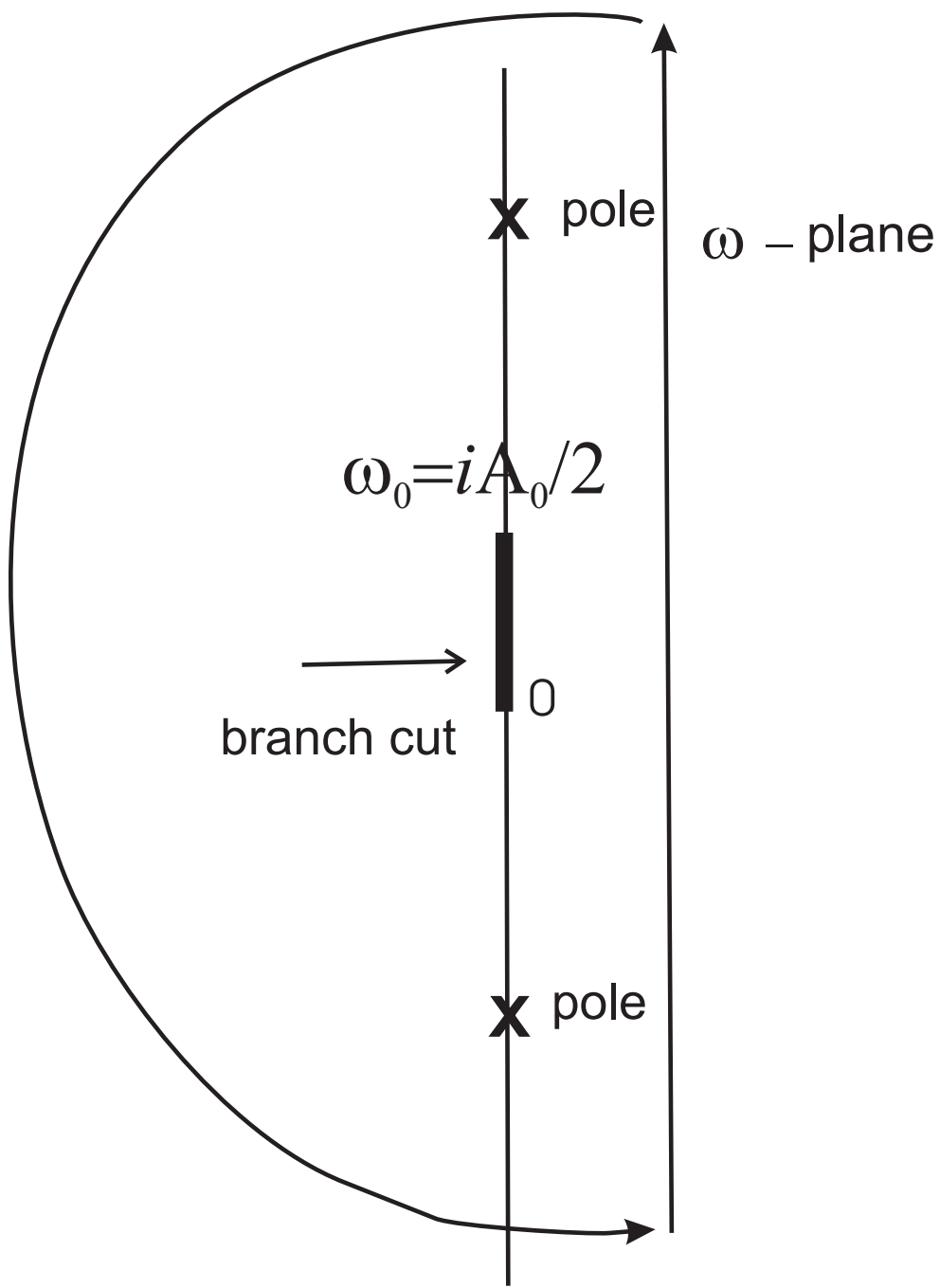}
\end{center}
\caption{The integration contour  $\Gamma$
 in Eq.(\ref{Zeeman}), enclosing poles and branch cut.}
\label{fig:3}
\end{figure}

\begin{figure}[h]
\begin{center}
\leavevmode
\includegraphics[width=13cm]{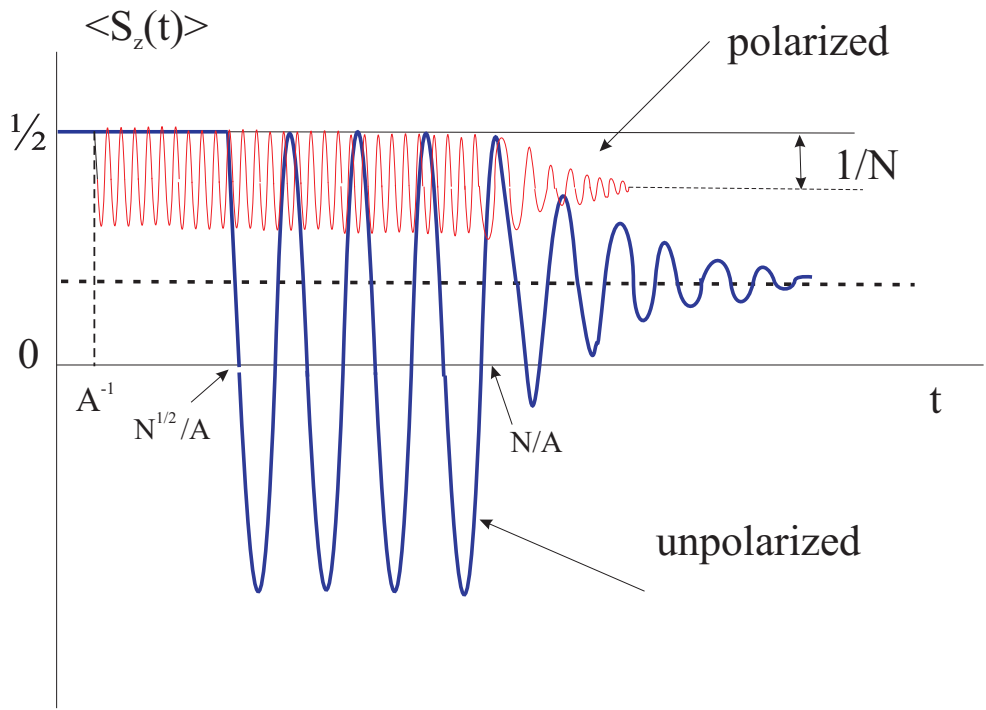}
\end{center}
\caption{Schematic  dependence of $<S_z(t)>$ on time t
for the unpolarized tensor product and fully polarized nuclear states. The time scale for the onset of the decay $\sim N/A$ is the same for both cases. In the fully polarized case the magnitude of the effect is $1/N$. The period of oscillations is of the order of $\sqrt{N}/A$ for the unpolarized and $\sim 1/A$ for the polarized case.}
\label{fig:4}
\end{figure}

\begin{figure}[h]
\begin{center}
\leavevmode
\includegraphics[width=13cm]{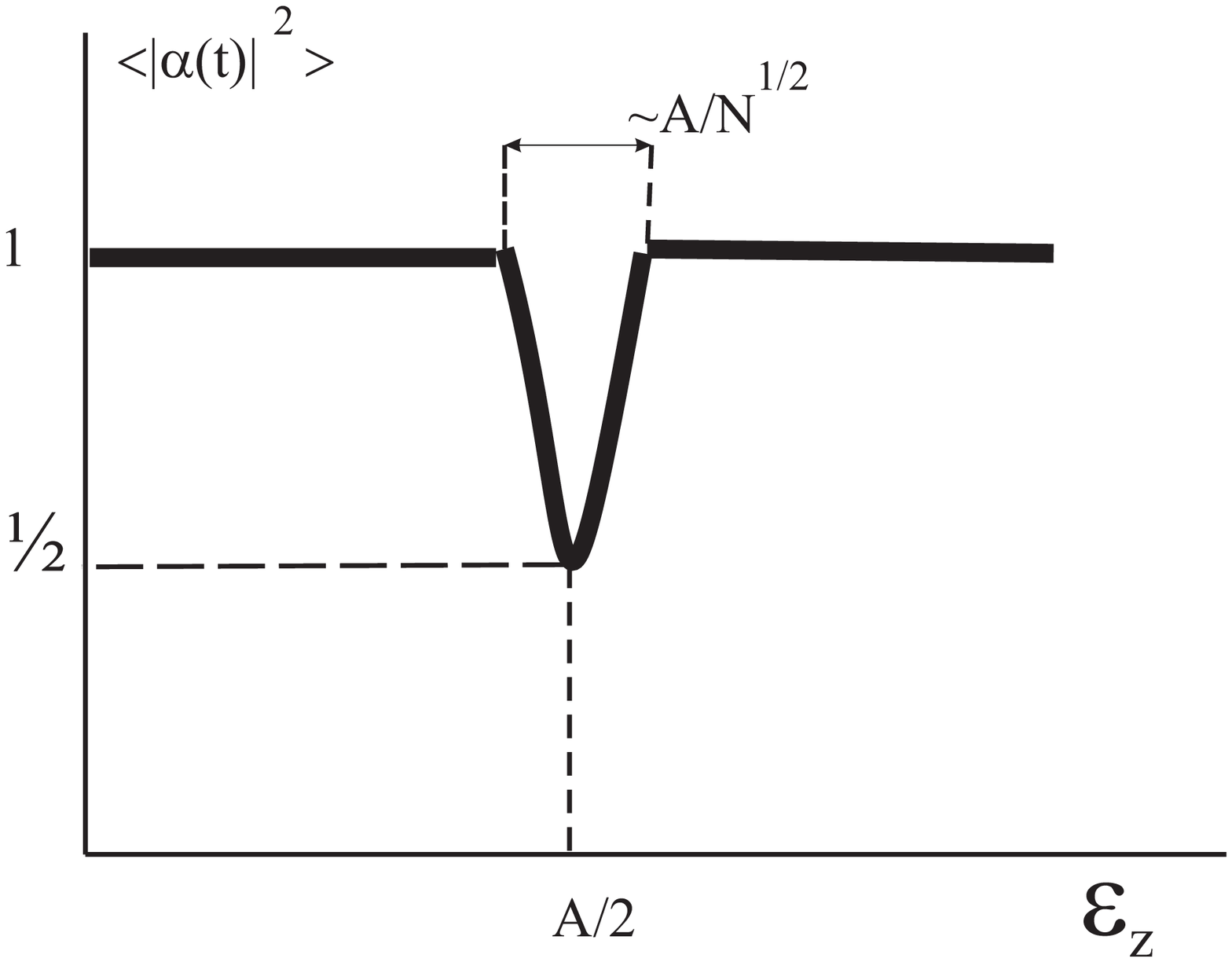}
\end{center}
\caption{The dependence of $|\alpha|^2$ averaged over time ($<|\alpha(t)|^2>$) on the external Zeeman field $\epsilon_z$ for a fully polarized nuclear state. The resonance occurs at $|\epsilon_z| = A/2$, and the width of the resonance is $\sim A/\sqrt{N}$, which is much smaller than the initial gap $A/2$.}
\label{fig:5}
\end{figure}


\begin{references}
\bibitem{ensemble} We note that in cases with exponential decay the spin decoherence
(dephasing) time is usually denoted by $T_2$ ($T_2^*$) (e.g. in the Bloch equations).
E.g., in Ref.\onlinecite{Awschalom} it is
$T_2^{\star}$ which is measured (due to an ensemble average over many spins). In
general, $T_2^{\star} < T_2$. 
Since the spin decay found in the present work turns out to be non-exponential, we will
not use this notation  (i.e. $T_2$ or $T_2^*$) to avoid confusion.
\bibitem{general}
Note that we consider special nuclear configurations $\{I_z^i\}$ which we can treat analytically. 
We stress, however, that  more general initial tensor product states which correspond to arbitrary directions of individual spins have the same time evolution as the state considered here, which was confirmed numerically in Ref.\onlinecite{Schliemann}. In contrast, a very different decay in time occurs for randomly entangled states \cite{Schliemann}. See also Sec.\ref{initial}.
\bibitem{conserv}
Since the dipolar interaction between the nuclei is neglected, the total angular momentum of the system ${\vec J} = {\vec S} + \sum_i {\vec I}_i$ is conserved, where the sum runs over the entire sample (and not only over the dot). However, the corresponding angular momentum obtained with the sum running over the nuclei located within the dot is not conserved, which allows a transfer of  nuclear moment to the  region outside the dot via the hyperfine interaction with the electron spin.
\bibitem{experim}
The nuclear spin correlator
$C(t)=\langle I^{z}(t)I^{z}(0)\rangle$, $\vec I=\sum_{i}\vec I_{i}$, 
can be measured directly by local NMR-type measurements such as magnetic resonance
force microscopy \cite{Suter02}. As pointed out in Ref.\onlinecite{Schliemann}, in a subspace of given $J^{z}$ and with the electron spin pointing downwards initially, this quantity reads
$C(t)=(J^{z}-\langle S^{z}(t)\rangle)(J^{z}+1/2)$. 
 Then the dynamics of the total nuclear spin can be mapped out by measuring the
electron spin and vice versa.
\end{references}
\end{document}